\documentclass[runningheads]{llncs}
\usepackage{graphicx}
\begin{document}
\title{Next-Day Bitcoin Price Forecast Based on Artificial intelligence Methods\thanks{This paper is only a draft (part of a study).}}
%
%
\author{Liping Yang\inst{1}}
\authorrunning{Y. et al.}
%
\institute{School of Economics and Management, Beijing University of Chemical Technology
\email{lipingphd@163.com}\\}
\maketitle              
\begin{abstract}

In recent years, Bitcoin price prediction has attracted the interest of researchers and investors. However, the accuracy of previous studies is not well enough. Machine learning and deep learning methods have been proved to have strong prediction ability in this area. This paper proposed a method combined with Ensemble Empirical Mode Decomposition (EEMD) and a deep learning method called long short-term memory (LSTM) to research the problem of next-day Bitcoin price forecast. The result show that there are still more work need to be down when using the method of EEMD-LSTM method.

\keywords{Bitcoin  \and  Long short-term memory \and Deep learning \and Prediction \and EEMD}
\end{abstract}
\section{Introduction}

The cryptocurrency market has boomed in the past decade. Bitcoin, the world’s first decentralized and currently biggest digital currency, was introduced in 2008 by a group of programmers using the pseudonym ‘Satoshi Nakamoto’ (Chen, Chang \& Cheng, 2019)[\ref{refe.1}]. According to the “coinmarket-cap” website, Bitcoin accounted for more than 48\% of the market value of cryptocurrency (Chen, Xu \& Jia, 2021)[\ref{refe.2}]. \\

Due to the innovation and market position of Bitcoin, many studies have investigated the price prediction problem. The different Bitcoin price prediction approaches can be divided into statistical models and artificial intelligence (AI) model. Statistical models such as the autoregressive integrated moving average (ARIMA) and exponentially weighted moving-average models are the most commonly used analytical tools  (Munim, Shakil \& Alon,2019; Nguyen \& Le, 2019, November)[\ref{refe.3}-\ref{refe.4}]. Although these approaches have been used widely, they can only deal with liner problems and the variables must follow a normal distribution. With the popularity of machine learning and deep learning methods (Jordan \& Mitchell, 2015;  Goodfellow, Bengio \& Courville, 2016)[\ref{refe.5}-\ref{refe.6}], relevant tools have also been used in the prediction research of Bitcoin price. Literature \ref{refe.2}(Chen, Xu \& Jia, 2021) used a two-stage model, and on the basis of obtaining relevant economic and technical features, random forest model (RF) and artificial neural network (ANN) were first used for feature extraction, and then use long short-term memory (LSTM) model to obtain predictions. In addition, Munim (2019)[\ref{refe.3}] analyzed forecasts of Bitcoin price using the ARIMA model and neural network autoregression (NNAR) models and demonstrates that ARIMA enduring power of volatile Bitcoin price prediction.\\

The main contributions of this paper are summarized as follows.

(1)	A data decomposition method called EEMD is used to distinguish high and low frequency components in order to fully explore the characteristics of each component.

(2)	A deep learning method called long short-term memory (LSTM) model was used to research the problem of next-Day Bitcoin price forecast.

The remainder of this paper is organized as follows. In Section 2, we explain the methods proposed by this paper. In Section 3, we describe the data used in this study and present our analyses and discuss the results. Finally, in Section 5, we give our conclusions.

\section{Material and Methods}
\subsection{Empirical mode decomposition}

The empirical mode decomposition (EMD) method is able to decompose a signal into some IMFs. An IMF is the function that satisfies the two following conditions: 

(1) in the whole data set, the number of extrema and the number of zero-crossings must either equal or differ at most by one.

(2) at any point, the mean value of the envelope defined by local maxima and the envelope defined by the local minima is zero (Huang, Shen \& Wu,1998)[\ref{refe.7}]. \\ 

An IMF represents simple oscillatory mode embedded in the signal. With the simple assumption that any signal consists of different simple IMFs, the EMD method was developed to decompose a signal into IMF components. The EMD process of a time series $x(t)$ can be described as follows:\\

(1).Initialize: $\gamma_0=x(t)$ and $i=1$

(2).Extract the $i$th IMF.

    (a).Initialize:$h_{i(k-1)}=r_i,k=1$.

(b).Extract the local maxima and minima of $h_{i(k-1)}$.

(c).Interpolate the local maxima and the minima by cubic spline lines to form upper and lower envelops of $h_{i(k-1)}$.

(d).Calculate the mean $m_{i(k-1)}$ of the upper and lower envelops of $h_{i(k-1)}$.

(e).Let $h_{i(k)}=h_{i(k-1)}-m_{i(k-1)}$

(f).If $h_{i(k)}$ is a IMF then set $IMF_i=h_{ik}, else go to step(2) with k=k+1$.

(3).Define $r_{i+1}=r_i-IMF_i$.

(4).If $r_{i+1}$ still has least 2 extrema then go to step(2) else decomposition process is finished and $r_{i+1}$ is the residue of the time series.\\

At the end of the procedure we have a residue $r_I$ and a collection of IMFs $c_i(i=1,2,...,I)$. Summing up all IMFs and the final residu $r_I$, we obtain:
\begin{equation}
    x(t)=\sum_{i=1}^{I}{c_i+r_I}
\end{equation}

Thus, we can achieve a decomposition of the signal into $I$ IMFs and a residue $r_I$, which is the mean trend of $x(t)$. The IMFs $c_1,c_2,...,c_I$ include different frequency bands ranging from high to low. The frequency components contained in each frequency band are different and they change with the variation of signal $x(t)$, while $r_I$ represents the central tendency of signal $x(t)$. A more detailed explanation of EMD can be found in Ref.\ref{refe.7}(Huang, Shen \& Wu,1998)\\

\subsection{Ensemble Empirical Mode Decomposition}

EEMD is an improved version of the EMD method (Wu \& Huang,2009)[\ref{refe.8}]. Compared with EMD, EEMD not only cancels out the added white noise, but also retains the signal information of the original time series, which makes the mode aliasing problem controlled to a certain extent. The specific steps are as follows:\\

1. Add a group of white noises to the original time series to form a signal-noise mixture;

2. The above signal-noise mixture with white noise was decomposed by EMD to obtain each IMF component;

3. Repeat steps 1 and 2, adding new white noise each time;

4. The IMF obtained each time will be integrated and averaged as the final results.

\subsection{Long short-term memory}

\begin{figure}
\includegraphics[width=0.55\textwidth]{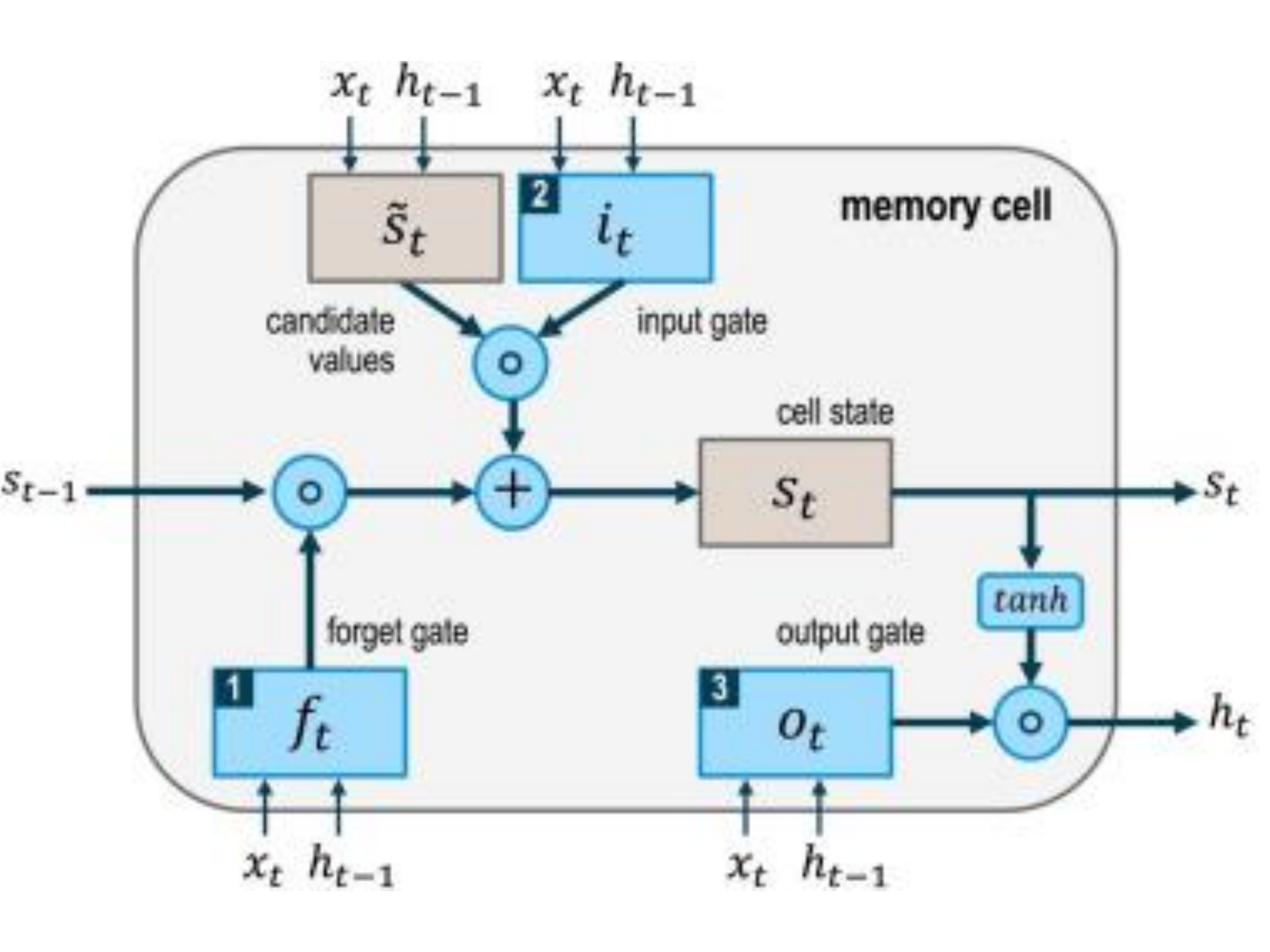}
\centering
\caption{Memory cell used by LSTM.} \label{fig1}
\end{figure}

The LSTM proposed by (Hochreiter \& Schmidhuber, 1997)[\ref{refe.9}] is the most commonly used model in recursive neural networks (RNNs). The LSTM can obtain better time series predictions because it is difficult to adapt many classical linear methods to multivariate or multi-input prediction problems. Compared with RNNs, the LSTM is characterized by the addition of memory cells for information processing. The structure of a memory cell is illustrated in Fig.\ref{fig1} (Chen, Xu \& Jia, 2021).

As shown in Fig.\ref{fig1}, at every timestep $t$, each of the memory cells has three gates, which maintain and adjust their cell state $S_t$: a forget gate ($f_t$), an input gate ($i_t$), and an output gate ($o_t$). Each gate serves a different purpose, as follows.\\

Input gate $i_t$ controls the input information flowing into the memory cell, where

\begin{equation}
    i_t=\sigma(W_{i,x}x_t+W_{i,h}h_{t-1}+b_i)
\end{equation}

Forget gate $f_t$ controls the forgetting of information in the cell, where

\begin{equation}
    f_t=\sigma(W_{f,x}x_t+W_{f,h}h_{t-1}+b_f)
\end{equation}

Output gate $o_t$ controls the output information flowing out of the cell, whre

\begin{equation}
    o_t=\sigma(W_{o,x}x_t+W_{o,h}h_{t-1}+b_o)
\end{equation}

The input features are calculated by intput $x_t$ and the previous hidden state $h_{t-1}$ by a tanh function, as follows:
\begin{equation}
    \tilde{s_t}=f_t\circ{s_{t-1}}+i_t\circ{\tilde{s_t}}
\end{equation}

The output of the LSTM at the time $t$ is then derived as:
\begin{equation}
    h_t=o_t\circ{tan(s_t)}
\end{equation}

where $\circ$ denotes the Hadamard product.

Finally, we project the output $h_t$ to the predicted output $\tilde{y_t}$ as:
\begin{equation}
    \tilde{y_t}=W_th_t
\end{equation}

Where $W_t$ is a projection matrix for reducing the dimension of $h_t$.\\

Due to the relatively samll sample size, we used a simple LSTM structure that included an input layer, hidden layer and output layer. The LSTM was implemented using the "keras" package in Python.

\section{Results}
\begin{figure}
\setlength{\belowcaptionskip}{-1.cm}
\includegraphics[width=0.85\textwidth]{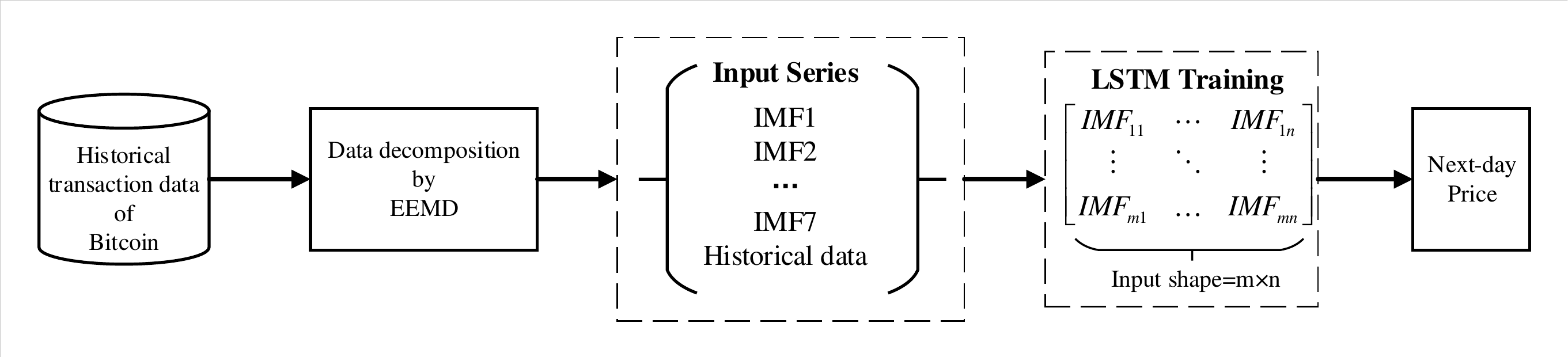}
\centering
\caption{The overall flow of methodology} \label{fig6}

\end{figure}

The general overview of the research methods adopted in this paper is shown in Fig.\ref{fig6}.
\subsection{ Data collection}

\begin{figure}
\setlength{\belowcaptionskip}{-1.cm}
\includegraphics[width=0.85\textwidth]{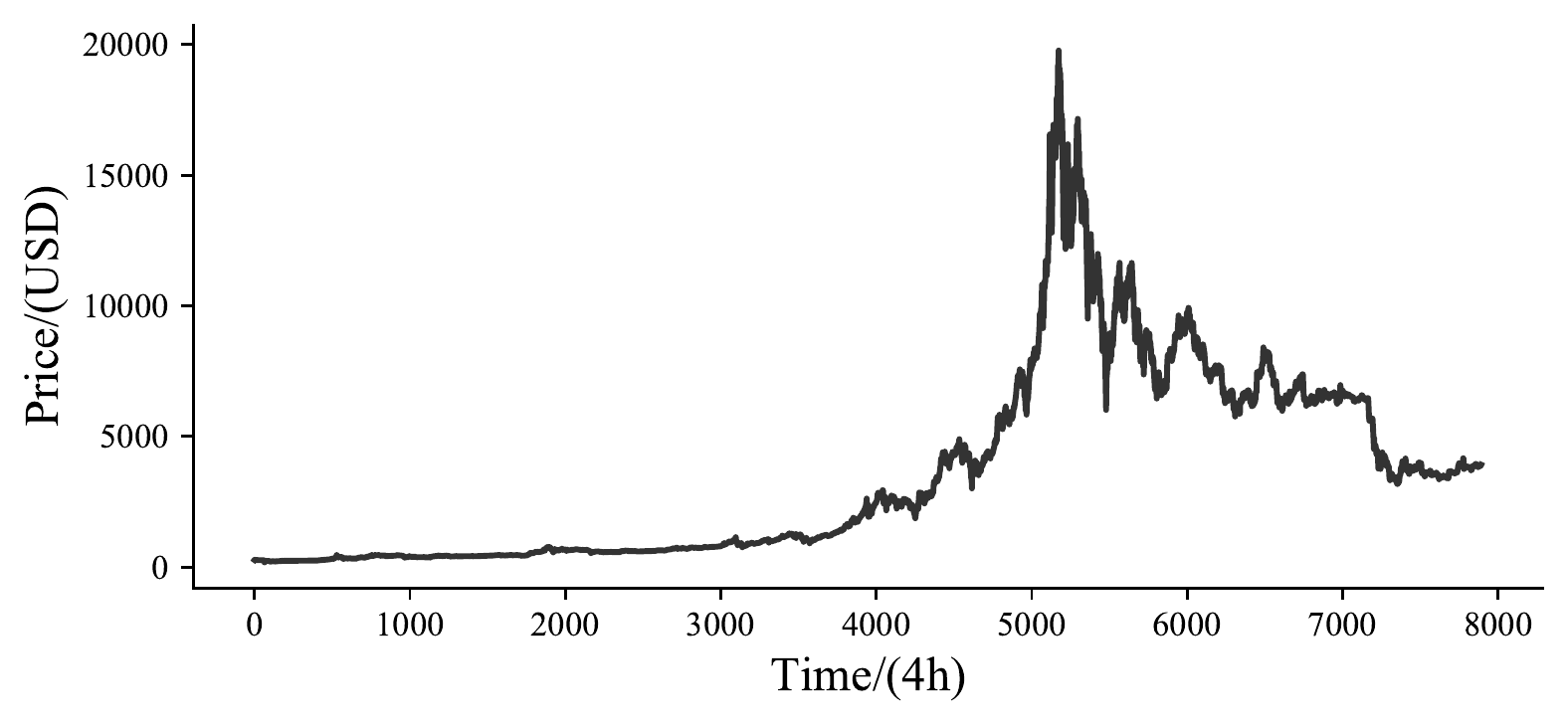}
\centering
\caption{Bitcoin trading price(4h)} \label{fig2}
\end{figure}

The data came from the standard data set of Bitcoin research published by the research team of Sun Yat-sen University in China (Han, Wu \& Zheng, 2020)[\ref{refe.10}]. This dataset is the market data of Bitcoin in terms of price and volume from August 2015 (when Ether first appeared) to March 2019. The time interval of sampling is selected as four-hour, that is to say, we choose every kinds of price and volume every of four-hour as the original data. The original market data of Bitcoin are obtained from Poloniex, one of the most active crypto asset exchanges.The data is shown in Fig.\ref{fig2}.

\subsection{Data decomposition and pre-processing}

\begin{figure}
\setlength{\belowcaptionskip}{-1.cm}
\includegraphics[width=0.65\textwidth]{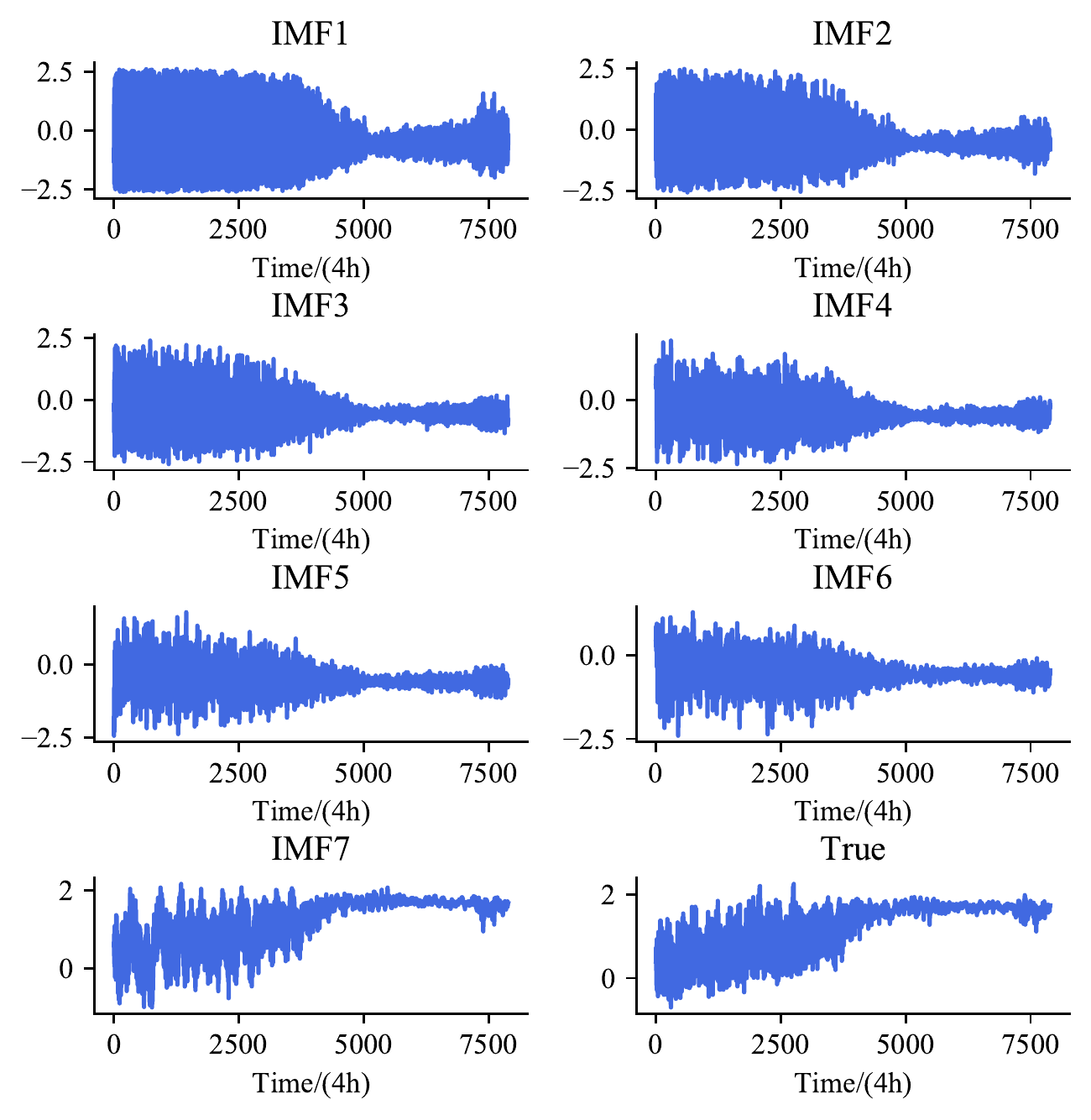}
\centering
\caption{The results of data decomposition and pre-processing} \label{fig3}
\end{figure}

The EEMD method was used to specify the white noise as 0.2 and the maximum IMF number as 7. A pre-processing method called scale is used for data pre-processing, as shown in Equation
\begin{equation}
    x=\frac{x-\overline{x}}{\sigma^2}
\end{equation}
$\overline{x}$ represents the numerical average value of the variable and $\sigma^2$ is the variance of this variable.The results of data decomposition and pre-processing are shown in the Fig.\ref{fig3}.

\subsection{Next-day prediction}
\begin{figure}
\includegraphics[width=0.65\textwidth]{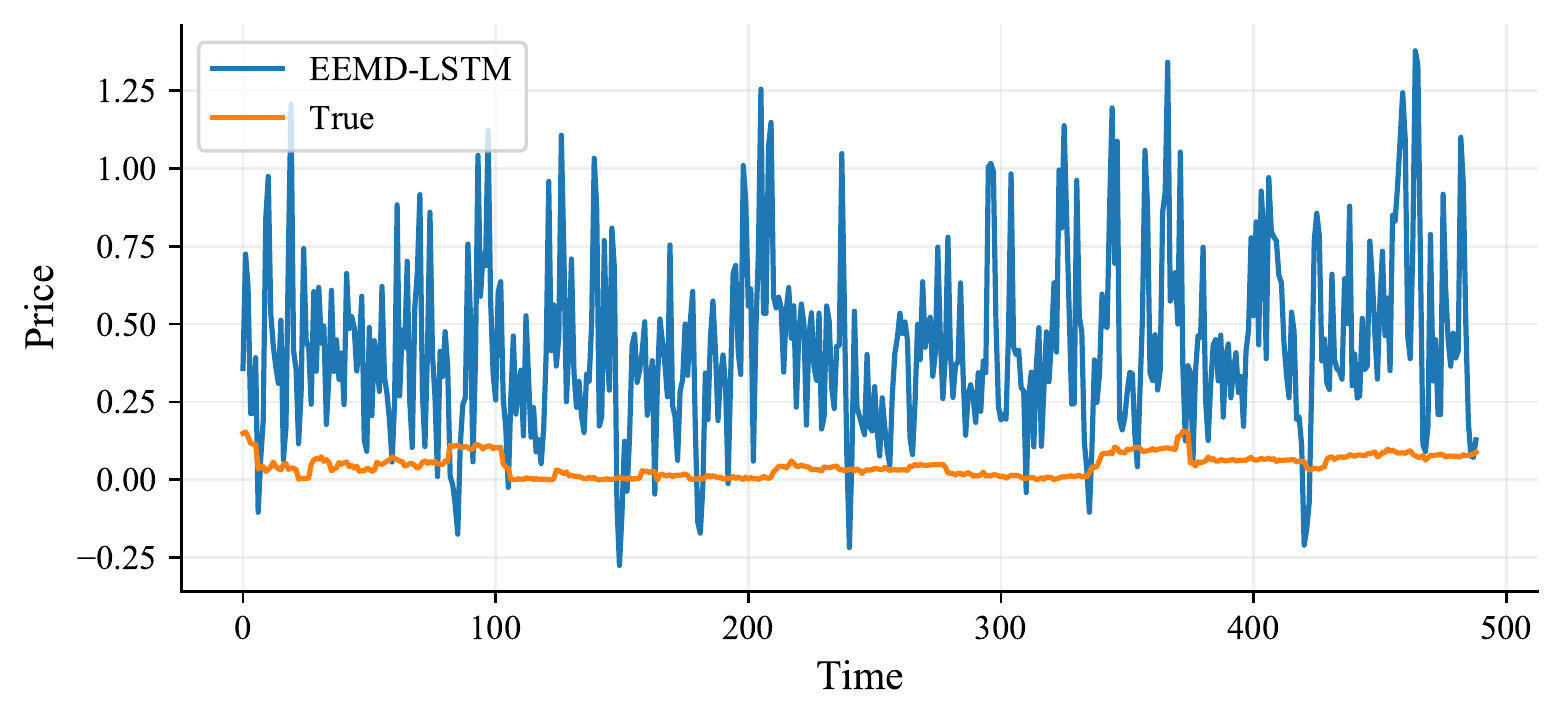}
\centering
\caption{Prediction results of EEMD-LSTM} \label{fig5}
\end{figure}

Use the rmsprop method as the optimization method and specify loss as MSE. 70\% of the data are used as the training set, 15\% as the verifier, and 15\% as the test set. LSTM. The specific design is as follows: step size is 3, input feature number is 7. The result of prediction is shown in Figure \ref{fig5}

\section{Conclusion}
The result showed that the EEMD-LSTM method is not good enough to predict the next-day price of Bitcoin. In the future, I will do more works about the theme, such as introduce more powerful models or re-train the EEMD-LSTM model.

\end{document}